\def\n{\noindent}

\def\P{$P$}

\def\eq{\enskip =\enskip}
\def\pls{\enskip +\enskip}
\def\mns{\enskip -\enskip}

  \def\ket{\vert \vert  \{ \emptyset \} \rangle}
  \def\ket2{\vert \vert \otimes \{ R \} \rangle}

\def\.#1{\mathaccent 95#1}
\def\^#1{\mathaccent 94 #1}
\def\~#1{\mathaccent "7E #1}

\def\equal{\enskip =\enskip}
\def\plus{\enskip +\enskip}

\def\eq{\enskip =\enskip}
\def\pls{\enskip +\enskip}
\def\mns{\enskip -\enskip}

  \def\ket{\vert \vert  \{ \emptyset \} \rangle}
  \def\ket2{\vert \vert \otimes \{ R \} \rangle}

\def\gt{\; > \;}
\def\lt{\: < \:}
\def\be{\begin{equation}}
\def\ee{\end{equation}}

\documentclass[12pt,epsfig]{iopart}
\usepackage{epsfig}
\begin{document}
\setcounter{page}{1}
\title{Study of Phase Stability in NiPt Systems}
\author{\bf Durga Paudyal, Tanusri Saha-Dasgupta and Abhijit Mookerjee}
\address { S.N. Bose National Centre for Basic Sciences,
JD Block, Sector 3, Salt Lake City, Kolkata 700098, India\\ 
email: dpaudyal@bose.res.in, tanusri@bose.res.in, abhijit@bose.res.in}
\date{\today}
\begin{abstract} 
We have studied the problem of phase stability in NiPt alloy system. 
We have used the augmented space recursion based on the TB-LMTO as the
method for studying the electronic structure of the alloys. In particular,
we have used the relativistic generalization of our earlier technique.
We note that, in order to predict the proper ground state structures and
energetics, in addition to relativistic effects, we have to take into account 
charge transfer effects with precision. 
\end{abstract}


\section{Introduction.}
There has been growing interest in the study of alloy phase ordering and segregation using
first principles techniques.
 In order to study these phenomena one needs a derivation of the configurational
energy for the alloy system. Different models have been proposed in which 
 the configurational energies are expressed in terms of effective
multi-site interactions, in particular effective pair
interactions \cite{kn:epi}. The analysis of alloy
ordering tendencies and phase stability reduces to the accurate
and reliable determination of effective pair interactions .
There are two different approaches of
obtaining the effective pair interactions.  One approach is
to start with electronic structure calculations of the total
energy of ordered super-structures of the alloy and
to invert these total energies to obtain the effective pair
interactions. This is the Connolly-Williams method \cite{kn:cw}. 
The other approach is to start from  the completely 
disordered  phase, set up a perturbation  in the form of
concentration fluctuations associated with an ordered phase and
study whether the alloy can sustain  such  a  perturbation.   This
includes approaches like the generalized perturbation  method (GPM)\cite{kn:gpm},  the
embedded cluster method (ECM)\cite{kn:ecm}. Most of the works on  the electronic 
structure  of  the disordered alloys have been  based  so far on 
the coherent potential approximation (CPA).  The
CPA being a single-site approximation cannot take into account the effect at a site
of its immediate environment. 
In an attempt to go beyond the single site approximation,
de Fontaine and his  group  followed  a  different  approach  of
direct configurational averaging (DCA) \cite{kn:dca}, without resorting  to
any kind of  single-site approximation. The  effective  pair  and
multi-site interactions were calculated directly in real space for
given configurations and the averaging was done in a brute force way 
by summing over different configurations. Invariably, the number of configurations
was finite and convergence of the results with increasing number of configurations
is yet to be available.
\vskip 0.2cm

\n Saha \etal \cite{kn:sdm} have introduced the augmented space recursion (ASR) based on
the augmented space formalism (ASF) first suggested by Mookerjee \cite{kn:mook} coupled
with the recursion method of Haydock \etal \cite{kn:hhk}. Within ASF the configuration 
averaging is carried out without having to resort to any single-site approximation. The recursion method
allows us to take into account the effect of the environment of a given site . Moreover, the
convergence of various physical quantities calculated through recursion with the number
of recursion steps and subsequent termination has been studied in great detail \cite{kn:sm}.
Among advantages of the ASR in going beyond the single-site approximation is the possibility 
of inclusion of local lattice distortions which is important in the case of alloys with large size mismatch 
between components as in the case of NiPt. In an earlier paper Saha and Mookerjee
\cite{kn:latdis} had discussed the effect of local lattice distortion on the electronic structures
of CuPd and CuBe alloys using the ASR. This allows the structure matrices to
randomly take values $S^{AA}_{LL'}, S^{AB}_{LL'}, S^{BA}_{LL'}$ or $S^{BB}_{LL'}$, depending on the
occupation of the sites $R$ and $R^\prime$.
The augmented space recursion 
coupled with orbital peeling technique \cite{kn:op} to evaluate small energy differences 
associated with band structure energies has been successfully used in past to describe the 
phase formation in alloys \cite{kn:tsdthesis}.
\vskip 0.2cm
\n In the present communication we focus on the application of this method for phase stability 
study in NiPt alloys. This system of alloys is of importance because of the possible need for relativistic 
corrections due to the heavy mass of Pt as well as effects due to  charge transfer and size mismatch between Ni and Pt. 
This therefore 
forms a perfect candidate for testing the applicability and limitations of our formalism, 
bringing in the relative importance of various effects for the accurate description of the 
system.
\noindent The previous studies of ordered and substitutionally disordered NiPt alloy 
systems have shown the importance of inclusion of relativistic effects. 
Treglia and Ducastelle \cite{kn:treglia} had shown that late transition metal alloys
should exhibit phase separating tendencies but they argue that the exceptional
ordering behavior of NiPt is due to the relativistic corrections.  
In a first principle study, Pinski \etal \cite{kn:pinski} found that the disordered fcc Ni$_{1-x}$Pt$_x$ alloy 
at $x = 0.5$, calculated by means of the single site KKR-CPA, becomes unstable at low temperatures, to
a perturbation by a $\langle {100}\rangle$ ordering wave and concluded that the
corresponding long range ordered state (LRO) i.e. the $L{1_0}$ structure should be the predicted ground state 
for which the large size mismatch between Ni and Pt plays the main role and the effect of relativity 
can be neglected. 
However, Lu \etal \cite{kn:lu} pointed out that a local ordering tendency
determined by perturbation analysis, doesn't necessarily predict the correct LRO ground state if the
size mismatch of the two elements is large, as is the case for Ni and Pt and concluded that relativity 
is the sole reason for long range order in NiPt. The work of Singh \etal \cite{kn:sin} demonstrated  
that the relativistic effects do stabilize the ordered structures over the disordered solid solution. 
Recently Ruban \etal \cite{kn:rs} have studied the problem of phase stability in 
NiPt alloy system based on ordered calculations with the inclusion of Madelung energy with multipole corrections. In this 
paper, we examine the relativistic treatment of the Hamiltonian and charge transfer and lattice relaxation effects on the 
electronic structure and phase stability of face-centered cubic NiPt system at $25\%$, $50\%$ and $75\%$ of concentration 
of Pt. As mentioned already, the augmented space recursion (ASR) technique, which we use here, is capable of taking into 
account environmental effects, effects of short range order and local lattice relaxation effects due to size mismatches.
To circumvent the problem of calculation of Madelung energy contribution for disordered system, 
we have used the appropriate effective atomic sphere radii for each of the constituents so that the spheres are neutral 
on the average and this has been done with precision at each concentration \cite{kn:kd}.
We have shown that without inclusion of relativistic effects the formation energy comes out to be positive which 
contradicts experimental results. With the scalar relativistic corrections, involving mass-velocity and Darwin terms, 
the formation energy comes out negative indicating that the relativistic effects play an important role in NiPt alloys 
in agreement with earlier studies. We find that the charge transfer effects have also an important role to play in deciding 
on the correct ground state structure, particularly when the concentration of Pt is high. Our study on transition 
temperatures based on a mean field theory could reproduce the qualitative experimental trends.

\section{Formalism}

\subsection{The Effective Pair Interactions}

We start from a completely disordered alloy. Each site $R$ has an occupation
variable $n_R$ associated with it. For a homogeneous perfect disorder
$\langle n_R\rangle = x$, where $x$ is the concentration of one of the
components of the alloy. In this homogeneously disordered system we now
introduce fluctuations in the occupation variable at each site : $\delta x_R =
n_R - x$. Expanding the total energy in this configuration about the
energy of the perfectly disordered state we get : 

\begin{equation}
E(x) \eq  E^{(0)}\plus \sum_{R=1}^{N} E_{R}^{(1)}\ \delta x_{R} \plus
\sum_{RR'=1}^{N} E_{RR'}^{(2)}\ \delta x_{R}\ \delta x_{R'} \plus \ldots
\label{eq:eq1}
\end{equation}

\n  The coefficients $E^{(0)}$  , $E_R^{(1)} \ \ldots $  are  the  effective
renormalized cluster interactions. 
$E^{(0)}$  is the energy of the averaged disordered
medium. 
The renormalized pair interactions $ E_{RR'}^{(2)}$ express the correlation
between two sites and are the most dominant quantities for the
analysis of phase stability.    We  will  retain  terms  up  to  pair
interactions  in  the  configuration  energy  expansion. Higher
order interactions may be included for a more accurate and
complete description. For the phase stability study it is the pair interaction which 
plays the dominant role. 
\vskip 0.2cm
\n The total energy of a solid may be separated into two terms : a
one-electron band contribution $E_{BS}$ and the electrostatic contribution $E_{ES}$
The renormalized cluster interactions defined in (~\ref{eq:eq1})
should, in principle, include both $E_{BS}$ and $E_{ES}$
contributions. Since the renormalized cluster interactions
involve the difference of cluster energies, it is usually assumed
that the electrostatic terms cancel out and only the band
structure contribution is important. Such an
assumption which is not rigorously true, has been shown to be
approximately valid in a number of alloy systems \cite{kn:heine}.
\n Considering only band structure contribution, the effective pair interactions may be written as :

\begin{equation} 
E_{RR'}^{(2)} \eq -\int_{-\infty}^{E_F} dE \left\{  -\frac{1}{\pi} \Im m~\log \sum_{IJ}\det \left(G^{IJ}\right)(E)~ \xi_{IJ}\right\} \end{equation}

\n where,
 $G^{IJ}$ represents the configurationally averaged Green function corresponding to the disordered Hamiltonian 
whose R and R$^\prime$ sites are occupied by I-th and J-th type of atom, and

\[ \xi_{IJ} \eq \left\{\begin{array}{ll}
                         +1  & \mbox{if I=J}\\
                         -1  & \mbox{if I$\not=$ J}
	                 \end{array} \right. \]

\n The behavior of this function is quite complicated and hence the integration by standard routines (e.g. 
Simpson's rule or Chebyshev polynomials) is difficult, involving many iterations before convergence is achieved. 
Furthermore the integrand is multi-valued, being simply the phase of $\sum_{IJ}\det \left(G^{IJ}\right)~\xi_{IJ}$. 
The way out for this was suggested by Burke \cite{kn:op}which relies on the repeated application of the 
partition theorem on the Hamiltonian H$^{IJ}$. The final result is given simply in terms
of the zeroes and poles of the Green function in the region $E<E_F$ 

\begin{equation}
\fl E^{(2)}_{RR'} \eq 2 \sum_{IJ}\xi_{IJ} \sum_{k=0}^{\ell max} \left[ \sum_{j=1}^{z^{k,IJ}}\ Z^{k,IJ}_j \mns \sum_{j=1}^{p^{k,IJ}}\ P^{k,
IJ}_j
\pls \left( p^{k,IJ} - z^{k,IJ}\right)\ E_F\right] \end{equation}

\n where $Z^{k,IJ}_j$ and $P^{k,IJ}_j$ are the zeros and poles of the peeled Green's function $G^{IJ}_{k}$ of disordered Hamiltonian with 
occupancy at sites R and R$^\prime$ by I and J of which first (k-1) rows and columns has been deleted. $p^{k,IJ}$ and $z^{k,IJ}$ are 
the number of poles and  zeroes in the energy region below $E_F$. The factor $2$ accounts for the spin degeneracy.

\subsection{The Augmented Space Recursion}

    As discussed in the previous section, the calculation of the
effective pair interaction  in  our  formalism  reduces  to  the
determination of the peeled configuration averaged  green  functions
$\langle$G$^{IJ}_k\rangle $.
 We  shall  employ  the  augmented
space recursion coupled with the linearized tight-binding  muffin
tin orbital method (TB-LMTO) introduced by  Andersen  and  Jepsen
\cite{kn:lmto} for a first principles determination  of  these  configuration
averaged quantities.   We shall take the  most  localized,  sparse
tight  binding  first order Hamiltonian  derived systematically from the
LMTO theory within the atomic sphere approximation (ASA) and generalized to random alloys.  The augmented space recursion method has been described at 
great length in earlier communications (\cite{kn:sdm}-\cite{kn:sm},
\cite{kn:tsdthesis},\cite{kn:asf},\cite{kn:big}). We refer the
readers to these references for the details. We shall give here
the final form of the effective Hamiltonian used for recursion in
augmented space for the calculation of the peeled Green functions :

\begin{eqnarray}
\fl H^{IJ}_{k}  =  \sum_{\ell =k}^{\ell max} C^{I}_{R,\ell}a^{\dag}_{R}a_{R} +
\sum_{\ell =1}^{\ell max} C^{J}_{R',\ell}\ a^{\dag}_{R'}a_{R'}  +
      \sum_{R''\not= R,R'} \sum_{\ell =1}^{\ell max} \left( C^{B}_{R'',\ell } + \delta C_{\ell }
         {\widetilde{\mathbf M}}^{R''} \right)\  a^{\dag}_{R''} a_{R''} + \ldots\nonumber \\
\fl     +  \sum_{R''\not= R}\ \sum_{L=k}\sum_{L^{\prime}} 
          \Delta^{1/2,I}_{R,\ell }
         S^{R,R''}_{LL^{\prime}} \left( \Delta^{1/2,B}_{R'',\ell '} +
         \delta \Delta^{1/2}_{\ell '} {\widetilde{\mathbf M}}^{R''} \right)\  a^{\dag}_{R}a_{R''}  + \ldots \nonumber \\
\fl     +  \sum_{R''\not= R'}\ \sum_{L}\sum_{L^{\prime}}
          \Delta^{1/2,I}_{R',\ell }
         S^{R',R''}_{LL^{\prime}} \left( \Delta^{1/2,B}_{R'',\ell '} +
         \delta \Delta^{1/2}_{\ell '} {\widetilde{\mathbf M}}^{R''} \right)\ a^{\dag}_{R'}a_{R''} + \ldots \nonumber \\
 \fl    +  \sum_{R''\not= R}\ \sum_{L}\sum_{L^{\prime}=k}\left( \Delta^{1/2,B}_{R'',\ell} +
         \delta \Delta^{1/2}_{\ell} {\widetilde{\mathbf M}}^{R''} \right) 
         S^{R'',R}_{LL^{\prime}}\ \Delta^{1/2,I}_{R,\ell '}\ a^{\dag}_{R''}a_{R}
         + \ldots \nonumber \\
\fl    +  \sum_{R''\not= R'}\ \sum_{L}\sum_{L^{\prime}}\left( \Delta^{1/2,B}_{R'',\ell} +
         \delta \Delta^{1/2}_{\ell } {\widetilde{\mathbf M}}^{R''} \right)
         S^{R'',R'}_{LL^{\prime}}\ \Delta^{1/2,I}_{R',\ell '}\ a^{\dag}_{R''}a_{R'}
         + \ldots \nonumber \\
\fl     +  \sum_{R''\not= R,R'}\ \sum_{R'''\not= R,R'}\ \sum_{L}
         \sum_{L^{\prime}}\ 
         \left( \Delta^{1/2,B}_{R'',\ell} + \delta\Delta^{1/2}_{\ell} {\widetilde{\mathbf M}}^{R''} \right)
         S^{R'',R'''}_{LL^{\prime}}\ \left( \Delta^{1/2,B}_{R''',\ell '} +
          \delta\Delta^{1/2}_{\ell '} {\widetilde{\mathbf M}}^{R'''} \right) \ldots\nonumber\\
\phantom{xxxxxxxxxxx}    \ldots\ (a^{\dag}_{R''}a_{R'''}+a^{\dag}_{R'''} a_{R''})
\label{ham}
\end{eqnarray}
here, L is a composite index ($lm$)

\n For a  binary  distribution
$\widetilde{\bf M}^{R}$   is given by:

\begin{equation}
\widetilde{\bf M}^{R} \equal x\ b^{\dagger}_{R\uparrow}b_{R\uparrow} +
(1-x)\ b^{\dagger}_{R\downarrow}b_{R\downarrow}
+ \sqrt{x(1-x)}\ \left( b^{\dagger}_{R\uparrow}b_{R\downarrow} + b^{\dagger}_{R\downarrow}b_{R\uparrow}
\right)  \label{eq:m}
\end{equation}

\n For non-isochoric alloys , the difference in
atomic radii of the constituents lead to change in the electronic
density of states, as confirmed by experiment \cite{kn:cupd} and
approximate theoretical techniques \cite{kn:cupd2}. One thus
expects that the mismatch of size produces, in addition to a
relaxation energy E$_{R}$ contribution, a change in the band
structure. Within our Augmented Space Recursion (ASR),
off-diagonal disorder in the structure matrix S$^{\beta}$ because
of local lattice distortions due to size mismatch of the
constituents, can be
handled on the same footing as diagonal disorder in the potential
parameters \cite{kn:big}.
 
\n         The augmented space recursion with the TB-LMTO  Hamiltonian
coupled with orbital peeling allows us to  compute  configuration
averaged  pair-potentials  directly,  without  resorting  to  any
direct averaging over a finite number of  configurations.  In an earlier communication
\cite{kn:mook} we have discussed how one uses the local symmetries of the
augmented space to reduce the Hamiltonian and carry out the recursion on
a reducible subspace of much lower rank. If we fix the occupation of two
sites, the local symmetry of the augmented space is lowered (this is very
similar to the lowering of spherical symmetry to cylindrical symmetry
when a preferred direction is introduced in an isotropic system). We
may then carry out the recursion in a suitably reduced subspace.

\subsection{Static concentration wave method}
The static concentration wave (SCW) was proposed as a  theory for ordering by Khachaturyan (\cite{kn:kha78},\cite{kn:kha83}).
The occupation probability $n(\vec{r})$ plays the key role
in this theory. 
This function $n(\vec{r})$ that determines the distribution of solute atoms in an
ordered phase can be represented as a superposition of concentration waves:
\begin{equation}
n(\vec{r}) = c + \frac{1}{2} \sum_{j} [Q(\vec {k_j}) \exp (i\vec {k_j}\cdot \vec{r})
 + Q^{*}(\vec {k_j}) \exp (-i\vec {k_j}\cdot \vec{r})]
\end{equation}
where $\exp (i\vec {k_j}\cdot \vec{r})$ is a static concentration wave, $\vec {k_j}$ is a non zero wave
vector defined in the first Brillouin zone of the disordered alloy, $\vec{r}$ is a site vector of
the lattice  $\{\vec{r}\}$, index $j$ denotes the wave vectors in the Brillouin zone, $Q(\vec {k_j})$
is static concentration wave amplitude and $c$ is the atomic fraction of the alloying element. \\
\vskip 0.2 cm
\noindent The study of phase stability requires accurate approximations to the configurational energy as well as 
the use of statistical models to obtain the configurational entropy. The configurational energy within the pair 
interaction can be represented in Fourier space as the product of the Fourier transform of the effective pair 
interaction V($\vec k$) and that of the pair correlation function Q($\vec k$):
\[ E \simeq \left(\frac {N}{2}\right) \sum_{\vec {k}} V(\vec k) Q(\vec k)\]
where N is the number of atoms. 
Minimization of E will naturally occur for states of order characterized by maxima in the Q($\vec k$) pair correlation 
spectrum located in the regions of the absolute minima of V($\vec k$). Consequently, much can be predicted 
about the types of ordering to be expected from a study of the shape of V($\vec k$), particularly from a search of its 
absolute minima (special points). At these points, 
\[ \left| \nabla _hV(h)\right| = 0 \] 
This was pointed out by Lifshitz \cite{kn:lali80, kn:krsm64} and Khachaturyan \cite{kn:kha78, kn:kha83}. Different types 
of ordered structures can be related directly to the minima of V($\vec k$). In other words, given the knowledge of concentration 
wave vectors, one can readily predict the most stable ordered structure of the system at low temperatures. This is comparable 
to the knowledge derived from the studies like those based on X-ray, electron and neutron diffraction.  A peak at the $\Gamma$ 
point, $\vec{k}$ = (000), indicates the phase separation, while a peak at the $\Gamma$ point, $\vec{k}$ = (100), in a fcc lattice 
suggests ordering. Peaks away from special points may correspond to the formation of long period superstructures.   
With in a simple mean field approximation, the instability can be obtained in the following way:
If we add the expression for dominant quadratic term in the average energy to that of the configurational entropy 
under the simple mean field approximation we obtain an expression for the free energy: 
\[ F = \sum_{i,j} V_{ij}^{2} (n_{i}-c) (n_{j}-c) + k_{B}T \sum_{i}[n_{i}\ln n_{i} + (1-n_{i}) \ln(1-n_{i})]\] 
where $n_{i}$ is the concentration of the species A at the i-th site and c is the average concentration of that species. 
If we define a configuration variable $\gamma_{i}^{0}$ as $\langle\delta n_i\rangle_0$ (the symbol $\langle \cdots \rangle_0$ 
denotes micro-canonical averaging), which is the variable relevant to the stability analysis, then the harmonic term in the 
Taylor expansion of the above free energy is  
\begin{equation}
F^{2}~~=~~{\frac{N}{{2}}}~~\sum_h \Gamma^{*}(\vec k(h)) F(\vec k(h))\Gamma(\vec k(h))
\end{equation}
where,
$\vec{ k}(h)=2\pi h_{\alpha }\vec{ b}_{\alpha }$ and $\Gamma(\vec k(h)) = {\cal F}~(n(\vec r) -c)$. 
The stability of a solid solution 
with respect to a small concentration wave of given wave vector $\vec k(h)$ is guaranteed as long as 
$F(\vec k(h))$ is positive definite. Instability sets in when $F(\vec k(h))$ vanishes i.e.
\begin{equation}
F(\vec k(h)) = k_B~T^{i} + V(\vec k(h))~c~(1-c) = 0
\end{equation}
$T^{i}$ being the temperature at which the instability sets in for the 
considered concentration wave. It appears from the above expression that under a simple mean field approximation 
the spinodal is always a parabola in the (t, c) phase diagram, symmetric about x = 0.5. It is the concentration 
dependence of the effective pair interactions which brings about the asymmetry.

\section{Computational details}
\subsection{Convergence of augmented space recursion} 
The effective pair potentials are calculated at the Fermi level so one needs to be very careful about the 
convergence of the Fermi energy  as well as that of the effective pair potentials.
In fact, errors can arise in the augmented space recursion 
because one can carry out only finite number of recursion steps and then terminate the continued fraction 
using available terminators. Also one chooses a large but finite part of the augmented space nearest neighbour map and 
ignores the part of the augmented space very far from the starting state. This is also a source of error.
\vskip 0.2cm
\n For finding out the Fermi energy accurately, we have used the energy dependent version of augmented space recursion.
In this version of ASR the defining Hamiltonian is recast into an energy dependent Hamiltonian having only diagonal disorder. 
We then choose a few seed points across the energy spectrum uniformly, carry out recursion on those points and spline fit the 
coefficients of recursion through out the whole spectrum. This enables us to carry out large number of recursion steps 
since the configuration space grows significantly less faster for diagonal as compared with off diagonal disorder. For details see 
ref \cite{kn:sdgthesis}
\vskip 0.2cm
\n We have checked  the convergence of Fermi energy and effective pair potentials with respect to 
recursion steps and the number of seed energy points taking the case of NiPt$_3$ system. 
We have found that the Fermi energy and effective 
pair potentials converge beyond seven recursion steps and thirty five seed energy points. In our all calculations 
reported in the following have been carried out with eight recursion steps 
and thirty five seed energy points. 

\subsection{Antiphase boundary energy}
 
Kanamori and Kakehasi \cite{kn:kanamori} used the method of geometrical inequalities which is capable of 
searching for ground structure. They considered the energy of the three dimensional 
Ising like model:
\begin{equation}
E_c = \sum_k V_k P_k
\end{equation}
where, $V_k$ is the interaction constant of the K-th nearest neighbour interaction and 
$P_k$ is the total number of $k^{th}$ neighboring pairs in the given configuration.
Defining the anti-phase boundary energy $\xi$ by
\begin{equation}
\xi = - V_2 + 4~V_3 - 4~V_4~,
\end{equation} 
the authors proved rigorously that for $\xi \gt 0$ $L{1_2}$ and $L{1_0}$ are the corresponding superstructures possible at 
concentration 25 $\%$ and 50 $\%$ while for $\xi \lt 0$, one has the $DO_{22}$ and $A_2B_2$ superstructures.
We have applied these conditions in our calculations to find out the relative stability between $DO_{22}$ and $L{1_2}$ 
structures in Ni$_3Pt$ and NiPt$_3$ that between $A_2B_2$ and $L{1_0}$ in NiPt.  

\subsection{Special-point ordering}
 
A wide range of phenomena related to order-disorder and magnetic transitions
can be explained using the symmetry properties of the pair potentials (V$_{ij}$). If a symmetry
element (rotation, rotation-inversion or mirror plane) of the space group in \textbf{k}-space is located at point \emph{h},
the vector representing the gradient $\nabla _hV(h)$ of an arbitrary
potential energy function V(\emph{h}) at that point must lie along or within
the symmetry element. If two or more symmetry elements intersect at point
\emph{h}, one must necessarily have
\begin{equation}
\left| \nabla _hV(h)\right| =0
\end{equation}
since a finite magnitude vector can not lie simultaneously in intersecting straight lines having only a 
point in common.
At these so-called special points, the potential energy function V(\emph{h})
represent an extremum regardless of the choice of the pair interaction
energies. Thus special points play an important role in the
search for lowest energy ordered structures. 
The points which differ by a vector of a reciprocal lattice are considered equivalent.
In the case of simple structures with a single atom per unit cell, it is
sufficient that two symmetry elements intersect at special points.
These special points are listed in Crystallographic tables. They
are always located at the surface of the Brillouin zone. The `star' of a
special point vector \textbf{k} is obtained by applying all the rotations
and rotation-inversion of the space group on the vector \textbf{k}. All these
vectors of a star are also considered equivalent.
The special points of the $fcc$ structure are located at the points $\Gamma $,
X, W and L of the Brillouin zone as shown in table 1.
\begin{table}
\caption{The special points and stars of the $fcc$ structure}
\begin{center}
\begin{tabular}{|c|l|c|c|}
\hline\hline
\textbf{k}-vector Star & Members & Brillouin zone points & Ordering structure
\\ \hline
& & & \\
$\left\langle 000\right\rangle $ & $\left[ 000\right] $ & $\Gamma $ & \\
& & & \\
$\left\langle 100\right\rangle $ & $\left[ 100\right] $ $\left[ 010\right] $
$\left[ 001\right] $& X & L1$_2$, L1$_0$ \\
& & & \\
$\left\langle 1\frac{1}{2}0\right\rangle$ &
$\left[ 1\frac{1}{2}0\right]$ $\left[ \frac{1}{2}01\right]$
$\left[01\frac{1}{2}\right]$ & W & A$_2$B$_2$,~DO$_{22}$ \\
& & & \\
& $\left[\overline{1}\overline{\frac{1}{2}}0\right]$
$\left[\overline{\frac{1}{2}}0\overline{1}\right]$
$\left[0\overline{1}\overline{\frac{1}{2}}0\right]$ & & \\
& & & \\
$\left\langle \frac 12\frac 12 \frac 12 \right\rangle$ &
$\left[ \frac 12\frac 12 \frac 12 \right]$
$\left[ \frac 12\overline{\frac 12} \overline{\frac 12} \right]$
& L & L1$_1$ \\
& & & \\
&$\left[ \overline{\frac 12}\frac 12 \overline{\frac 12} \right]$
$\left[ \overline{\frac 12} \overline{\frac 12}\frac 12 \right]$
&  &  \\
& & & \\ \hline \hline
\end{tabular}
\end{center}
\end{table}
\subsection{Ordering energy}
The ordering energy is defined as the difference between the formation energy of ordered alloy and the corresponding
formation energy of disordered alloy. Since we are dealing with the effective pair potentials,
the ordering energy can be calculated using these pair
potentials. The relation for ordering energy using pair potentials is given as :
\begin{equation}
E^{ord} = \frac{1}{2} \sum_k V_k \delta x_o \delta x_k
\end{equation}
where, $\delta x_o~(\delta x_k) = x_o~(x_k) - x$, $x_o~(x_k) = 1$ if the site $o/k$ is occupied by A atom and $x_o=0$ 
if the site $o/k$ is
occupied by B atom. 
For $L{1_2}$ structure (for $Ni_3Pt$ and $NiPt_3$ in our case) the expression for ordering energy
per atom in terms of pair potentials considering only up to fourth nearest neighbours is given as:
\begin{equation}
E_{Ni_3Pt}^{ord} =  -\frac{3}{32} [V_1 -\frac{1}{3}~ V_2 + V_3
-\frac{1}{3}~V_4 ]
\end{equation}
For $L{1_0}$ structure for NiPt the expression for ordering energy
per atom considering up to fourth nearest neighbour pair potentials is given as:
\begin{equation}
E_{NiPt}^{ord} = -\frac{1}{8}  [V_1 - V_2 + V_3 - V_4]
\end{equation}
Using these two relations we have found the ordering energy for Ni$_3$Pt, NiPt and NiPt$_3$.

\section{Results and discussions}

\begin{table}[t]
\caption{The calculated equilibrium lattice parameters with the choice of neutral charge spheres including scalar 
relativistic corrections. The corresponding lattice parameters with out relativistic 
corrections are given in brackets.}
\begin{center}
\begin{tabular}{|c|c|}
\hline\hline
concentration     & eq. lattice parameter             \\
of Pt             &  in au. SR(NR)                    \\  \hline
0.00(Ni)          &  6.528(6.568)  (expt. 6.66)       \\
0.25(Ni$_3$Pt)    &  6.758(6.890)                     \\
0.50(NiPt)        &  7.127(7.335)                     \\
0.75(NiPt$_3$)    &  7.196(7.467)                     \\
1.00(Pt)          &  7.3685(7.683) (expt. 7.41)\\
\hline
\end{tabular}
\end{center}
\end{table}

We have applied our formalism discussed in the previous section in calculating the effective pair potentials for 
the fcc based NiPt alloys for concentrations $x = 0.25, 0.5$ and $0.75$ of Pt. The calculation of the effective 
pair potentials has been restricted up to fourth nearest neighbour interactions. Total energy density functional 
calculations were performed at the concentration $x = 0.25, 0.5$ and $0.75$ of Pt. 
The Kohn-Sham equations were solved in the local density approximation (LDA). The LDA was treated with in the context 
of linear muffin tin orbitals in the atomic sphere approximation. The calculations were performed non relativistically 
as well as scalar relativistically and the exchange correlation potential of Von Barth and Hedin was used. Two sets of 
calculations were performed one with the same Wigner-Seitz radius (charged spheres) for Ni and Pt. In other set 
we followed the procedure described by Kudrnovsk\'y \etal \cite{kn:kd} and used an extension of the procedure proposed 
by Andersen \etal \cite{kn:lmto}, which allows us flexibility in the choice of ASA radii for the constituents. The idea is to choose 
ASA radii of atomic species in such a way that the spheres are charge neutral on the average. The potential parameters $\Delta^{I}_l$ 
and $\gamma^{I}_l$ of the constituent I were then scaled by the factors $(s^{I}/s^{alloy})^{2l+1}$ to account for the fact 
that the Wigner-Seitz radius of constituent I, $s^I$, is different from that of the alloy, $s^{alloy}$. These potential 
parameters were used to parameterize the alloy Hamiltonian. For the purpose of augmented space recursion, seven shell 
map was generated and thirty five seed energy point recursion was performed, as explained in previous section, to calculate 
the Fermi energy with the second order LMTO-ASA Hamiltonian through the recursion method using eight level of 
recursion and analytical terminator of Luchini and Nex. For the effective pair potentials, we used the 
orbital peeling method within the frame work of ASR for the calculation of peeled averaged Green function 
described in detail in the earlier section.
\vskip 0.2cm

\begin{table}[t]
\caption{Formation energies for Ni$_x$Pt$_y$with the choice of neutral charge spheres including scalar relativistic correction. The values in brackets 
are without relativistic correction. The corresponding estimate for charged sphere 
calculations are shown with $*$'s. $**$ refers to calculations with out combined correction. $***$ refers to disordered formation energy}
\begin{center}
\begin{tabular}{|c|c|c|c|c|c|c|c|}
\hline\hline
$y$ & This work             &Expt.\cite{kn:alss}&FPLMTO+           &LMTO &LMTO+           &KKR-ASA\\
        & SR(NR)     &                   &CWM\cite{kn:alss}&   \cite{kn:alss}               &CWM\cite{kn:ruban2}  &(KKR-CPA)\\ \hline  
& & & & & &  \\
0.25         & -7.50(4.25)     &-5.16              &-6.30            &-7.17             &-6.66                &        \\
             & -7.59$^*$(4.17)$^*$  &                   &                 &                  &                  &        \\ \hline
& & & & & &  \\
0.50         & -9.44(4.74)     &-7.06              &-8.69            &-8.5              &-8.95                &-12.00$^{**}$ \cite{kn:sin}  \\ 
             & -9.02$^*$(4.85)$^*$  &                   &                 &                  &                  &-8.10 \cite{kn:sin2}       \\       
 
             &                    &                   &                 &                  &                  &(-7.7$^{***}$)\cite{kn:sin}    \\ \hline
 & & & & & &  \\
0.75         & -8.15(4.22)     &-4.78              &-6.40            &-6.70             &-9.12                &        \\
             & -3.97$^*$(6.65)$^*$  &                   &                 &                  &                     &        \\
\hline
\end{tabular}
\end{center}
\end{table}

\n In table 2, we have quoted the equilibrium lattice parameters that were used in our calculations. We obtained these
by minimizing the total energies with respect to the lattice parameters. 
 We have got slightly lower
equilibrium lattice parameters as compared to experimental ones. This is characteristic of the 
local density approximation which overestimates bonding.

\n In figure 1 we have shown the formation energy of NiPt alloy system with various Pt concentrations based on ordered 
calculations.
It shows that without inclusion of relativistic effects the formation energy
comes out to be positive which contradicts  experimental results. With the inclusion of scalar relativistic 
corrections the formation energy comes out to be negative. This indicates that relativistic effects play an important 
role in the stability of NiPt alloys, in agreement with earlier studies. 
Our results are in closer agreement with previous works based on the Full-Potential LMTO and the Connolly-Williams
technique (\cite{kn:alss},\cite{kn:ruban2},\cite{kn:fon94}) and with experimental estimate. 
Singh \etal \cite{kn:sin} have also calculated the formation 
energy for 50$\%$ of Pt. Their results for the formation energy obtained from ordered calculations without
combined correction deviates quite a bit from ours as well as other results based on the Full-Potential LMTO and the Connolly-Williams
technique (\cite{kn:alss},\cite{kn:ruban2},\cite{kn:fon94}), which is presumably due to the neglect
of the combined correction in reference \cite{kn:sin}.
Singh \etal \cite{kn:sin2} have also done the calculation with combined correction which shows better 
agreement. 
The full-potential methods are expected to provide better estimates than other
methods. 

\vskip 0.2cm

\begin{table}[p]
\caption{The effective pair potentials for NiPt alloy system calculated with 
potential parameters taken from calculations with the choice of charged spheres and including scalar relativistic correction.
(O-L) refers to calculations without multipole corrections, M refers to calculations with multipole
corrections and SCI to calculations with screened Coulomb interactions. US-PP refers to ultrasoft pseudo-potentials.
* refers to non-relativistic calculations.}
\begin{center}
\begin{tabular}{|c|c|c|c|c|}
\hline\hline
Reference & v$_1$         & v$_2$            &  v$_3$     & v$_4$     \\
         &  (mRy/atom)   & (mRy/atom)       & (mRy/atom) &(mRy/atom) \\    \hline
\multicolumn{5}{|c|}{\bf Concentration of Pt = 25\%}\\
\hline
Present work          &11.36          &-0.05             &-0.07       &-0.41      \\
                  &11.972$*$      &0.015$*$          &0.054$*$    &0.046$*$   \\
\hline
\multicolumn{5}{|c|}{\bf Concentration of Pt = 50\%}\\ 
\hline
Present work          &7.832          &0.114             &-0.129      &-0.057     \\ 
                  &8.597$*$       &0.10$*$           &0.053$*$    &0.263$*$   \\
\cline{1-1} & & & &   \\
Singh \etal \cite{kn:sin}    &4.22           &1.14              & 0.22       &-1.04      \\
               & 4.94*         & 0.52*& 0.32*& -0.18* \\
\cline{1-1} & & & &   \\
Pinski \etal \cite{kn:pinski}& 9.4* & 0.8* & 0.4*& -0.2*  \\
\cline{1-1} & & & &   \\
Pourovskii \etal \cite{kn:Pourovski}&             &           &            &             \\
CWM-ASA+M      & 5.00        & 0.25      & 0.19       & -0.28       \\
SGPM           & 5.28        & 0.06      & -0.82      & -0.66       \\
\cline{1-1} & & & &   \\
Ruban \etal \cite{kn:rs}    &             &           &            &             \\
with SGPM      &             &           &            &             \\
ASA+M (O-L)(SCI)& 14.05(15.44)&0.32(-0.10)&-1.09(-1.22)& -1.76(-0.84) \\
ASA        (SCI)& 12.26(14.35)&0.53(-0.15)&-1.31(-1.48)& -2.14(-0.98) \\
With Connolly Williams&       &           &            &               \\
ASA+M          & 12.68       &1.31       &-0.02       & -0.73         \\
ASA+M (O-L)    & 13.70       &0.49       &-0.86       & -1.39         \\
ASA            & 14.33       &0.28       &-1.72       & -1.92         \\
US(PP)         & 12.81       &1.30       &0.69        & -0.40         \\
Direct calculation(SCI)&     &           &            &               \\
ASA+M          & 12.45       &0.47       &-0.49       & -0.65         \\
\hline
\multicolumn{5}{|c|}{\bf Concentration of Pt = 75\%}\\
\hline
Present work           &2.785        &0.236      &-0.116      & 0.276         \\
                   &3.813$*$     &0.361$*$   &-0.175$*$   & 0.366$*$      \\
\hline
\end{tabular}
\end{center}
\end{table}

\begin{table}
\caption{The effective pair potentials for NiPt alloys with potential parameters 
taken from calculations with the choice of charge neutral spheres including scalar relativistic corrections. 
The corresponding estimate for non relativistic calculations are shown with $*$'s.}
\begin{center}
\begin{tabular}{|c|c|c|c|c|}
\hline\hline
Reference  & v$_1$       & v$_2$     &  v$_3$     & v$_4$       \\ 
          & (mRy/atom)  & (mRy/atom)&  (mRy/atom)& (mRy/atom)  \\    \hline  
\multicolumn{5}{|c|}{\bf Concentration of Pt = 25\%}\\
\hline
Present work           & 12.34       &-0.092     &-0.046      & -0.54       \\
               & 13.08$*$    &-0.021$*$  & 0.152$*$   & -0.041$*$   \\ 
\hline
\multicolumn{5}{|c|}{\bf Concentration of Pt = 50\%}\\
\hline
Present work           & 10.08       & 0.1       & 0.004      & -0.24       \\ 
               & 10.111$*$   & 0.126$*$  & 0.246$*$   &  0.175$*$   \\
\cline{1-1} & & & & \\
Singh \etal \cite{kn:sin}     & 16.02       & 1.34      & 0.06       & -1.58       \\
               & 11.96$*$    & 0.66$*$   & 0.28$*$    & -0.46$*$    \\         
\cline{1-1} & & & &   \\
Ruban \etal \cite{kn:rs}   &             &           &            &             \\
Neutral(GPM)   & 5.49        &1.22       &0.01        & -0.73        \\
\hline
\multicolumn{5}{|c|}{\bf Concentration of Pt = 75\%}\\
\hline
Present work           & 8.9         & 0.26      & 0.1        &  0.02         \\
               & 7.874$*$    & 0.297$*$  & 0.276$*$   &  0.34$*$      \\ \hline
\end{tabular}
\end{center}
\end{table}

\n We next approached the problem from the disordered end. We started from a completely disordered alloy and set up
concentration wave fluctuations in it to see when this destabilizes the disordered phase as suggested by
Khachaturyan \cite{kn:kha78}. The calculation of the lattice distortion for disordered alloys has been carried out 
within the structural model given by rigid ion structure (RIS) \cite{kn:dis}. According to this model the lattice 
relaxes in such a way as to keep all the nearest neighbour distances close to the sum of the corresponding atomic 
radii for a particular concentration. This 
is found to be a reasonable model to deal with lattice relaxation effects in non-isochoric alloys \cite{kn:latdis}. Due 
to the distortion of the lattice, the structure matrix S$_{LL^{\prime}}^{RR^{\prime}}$ 
(which is 9$\times$9 matrix for each ${RR^{\prime}}$ 
pair and for a spd basis set) can randomly take values S$_{LL^{\prime}}^{AA}$, S$_{LL^{\prime}}^{BB}$ and 
S$_{LL^{\prime}}^{AB}$ depending upon the occupying of 
sites R and R$^{\prime}$. 
\[ 
S^{RR^{\prime}}_{LL^{\prime}} = S^{AA}_{LL^{\prime}} n_R n_{R^{\prime}} + S^{AB}_{LL^{\prime}} 
[n_R (1-n_{R^{\prime}}) + 
(1-n_R) n_{R^{\prime}}] + S^{BB}_{LL^{\prime}} (1-n_R) (1-n_{R^{\prime}}) 
\]
where 
\[ n_{R} \eq \left\{\begin{array}{ll}
                         1  & \mbox{if R is occupied by A}\\
                         0  & \mbox{if R is occupied by B}
                         \end{array} \right. \] 
Considering the example of calculation of S$^{AB}_{LL\prime}$ where B is the larger atom  
({\it e.g.} Pt in the present case), this matrix for a specific pair among 12 nearest neighbours 
connects an A atom at the site (0, 0, 0) and a B atom, which in the undistorted 
case would have been at the position ($\frac{a}{2}~,\frac{a}{2}~,0$) is now at ($(\frac{a}{2}+d),~(\frac{a}{2}+d),~d$), 
where $d$ is the displacement 
due to lattice distortion and $a$ is the lattice constant. 
We have assumed that the lattice expands equally in the x, y and z directions. With these new coordinates and assuming that 
all other neighbouring coordinates are fixed at undistorted fcc positions (which is the essence of terminal point approximation 
\cite{kn:kd} ), we have computed the structure matrices $S^{AA}_{LL\prime}$, $S^{AB}_{LL\prime}$ and $S^{BB}_{LL\prime}$.  
This takes into account both the effect of radial distortion as well as angular distortion (the nearest neighbour 
is now $\sqrt{\frac{a^{2}}{2}+2ad+3d^2}$ 
instead of $\frac{a}{\sqrt{2}}$ and the nearest neighbour vector is ($(\frac{a}{2}+d)~,(\frac{a}{2}+d)~,d$) 
instead of ($\frac{a}{2}~,\frac{a}{2}~,0$) in the above example). The values of $d$ for S$^{AB}_{LL\prime}$ 
came out to be 0.064~$a$, 0.052~$a$ 
and 0.054~$a$ for 25$\%$, 50$\%$ and 75$\%$ concentration of Pt. 
The details of the calculation scheme can be found in reference \cite{kn:latdis}. In figure 2 we have shown the relative magnitudes 
of nearest neighbour distances for different concentrations of Pt in NiPt alloy system compared to Vegard's law values for average 
bond length.  
We have computed the effective 
pair potentials for two sets of potential parameters with charged and charge neutral spheres. Figure 3 shows that the effective 
pair potentials for NiPt$_3$ is very small in magnitude using potential parameters with charged spheres. 
We even used these pair potentials and calculated the anti-phase boundary energy according to the prescription 
described in the previous section. The anti-phase boundary energy comes out to be negative for NiPt$_3$ and NiPt 
indicating stability of DO$_{22}$ over L1$_2$ for NiPt$_3$ and A$_2$B$_2$ over  L1$_0$ for NiPt. Further we calculated the minima 
of the special points according to the prescription described in the previous section. In the case of NiPt$_3$ and NiPt 
shown in figure 5, we could not get the minima at $\langle 100 \rangle$ 
which is not quite correct because experiments show NiPt$_3$ has $L1_2$ and NiPt has $L1_0$ ordering. 
But in the case of Ni$_3$Pt we could get the positive anti-phase boundary energy as well as minima at $\langle 100 \rangle$ correctly 
showing the ordering $L1_2$. In figure 4 we have plotted the effective pair potentials as a function of 
energy relative to Fermi energy and number of neighboring shells with charge neutral potential parameters including scalar 
relativistic correction 
which shows that the first nearest neighbor pair potentials are larger in magnitude than 
the second, third and fourth nearest neighbour pair potentials.
With potential parameters from neutral sphere calculations including scalar relativistic
correction for NiPt$_3$ and NiPt the anti-phase boundary energies come 
out to be positive and the minima of special points are at $\langle 100 \rangle$ 
correctly showing $L1_2$ and $L1_0$ orderings. If we use 
charge neutral potential parameters with out including scalar relativistic effect the anti-phase boundary energies
come out to be positive for Ni$_3$Pt and NiPt but negative for NiPt$_3$. This shows for NiPt$_3$ both scalar relativistic 
as well as charge transfer effects play important role to predict correct ground state.  

\begin{table}
\caption{The anti-phase boundary energies for Ni$_x$Pt$_y$ alloys 
from charged and neutral sphere calculations}
\begin{center}
\begin{tabular}{|c|c|c|}
\hline\hline
concentration  & \multicolumn{2}{c|}{APB energy (mRy/atom)}      \\ \cline{2-3}
of Pt          & Charged spheres & Neutral spheres      \\ \cline{2-3}   
               & \multicolumn{2}{c|}{Relativistic(Non-Relativistic)}\\ \hline
  & & \\
0.25           & 1.41(0.017) & 2.07(0.793)   \\
0.50           & -0.402(-0.94)&0.876(0.158)          \\
0.75           & -1.804(-2.525)& 0.06(-0.553)         \\
  & & \\\hline
\end{tabular}
\end{center}
\end{table}

\n So, we argue that on increasing the concentration 
of Pt atom the careful treatment to take into account of charge transfer effect becomes increasingly important.
In figure 4, we have also shown the effective pair potentials 
without scalar relativistic 
corrections. For NiPt$_3$ it is clearly seen that the effective pair potentials with scalar 
relativistic correction are larger in magnitude than the non relativistic ones which is expected 
because of higher concentration of Pt.    
\vskip 0.2cm
\n In figure 6 we have plotted  that the effective pair potentials vs concentration of Pt with charge neutral potential 
parameters including scalar relativistic correction which shows that the first 
nearest neighbour effective pair potentials decrease with the increase of the Pt concentration. 
Singh \etal \cite{kn:sin, kn:singh} have also 
calculated the effective pair potentials using KKR-CPA-GPM method. Their values of effective pair potentials are much larger 
than ours. They pointed out that due to the large values of effective pair potentials
the ordering energy and ordering temperatures (transition temperatures) are much higher than that observed experimentally. 
Our estimates give rise to instability temperatures which are closer to the experimental results (shown in figure 7). For
example, our estimate for the instability temperature for the 50\% alloy is  1683$^o$K, whereas the estimate from the
KKR-CPA is around 2979$^o$K. The experimental estimates of the transition temperature is  950$^o$K \cite {kn:dahmani}.
In KKR-CPA-GPM method one considers only the single site approximation and one does not take into account any off diagonal disorder 
which may arise because of size mismatch of the constituent atoms. The ASR, on the other hand, as discussed earlier can
do this with facility. Our test calculation for NiPt (50\% concentration of Pt) without taking into account lattice relaxation 
due to size mismatch effect gives an estimate of instability temperature of 2363$^{0}$K which is indeed higher than that of our 
original estimate with taking into account lattice relaxation due to size mismatch effect. 
Furthermore, Singh \etal \cite{kn:sin, kn:singh} in their calculation for 
charge neutrality have taken the ratio of Wigner Seitz radii of Ni and Pt as 0.95. We, on the other hand, have
varied the ratio, with the provision that the total volume is conserved, till, on the average, the spheres
become charge neutral. We have observed the ratio to be 0.909, 0.913 and 0.919 for the Ni$_3$Pt, NiPt and NiPt$_3$.
Given these calculational differences, it is not surprising that our calculations result 
in smaller values of the pair-potentials
leading to better estimates of the instability temperatures. The calculations of Pinski \etal \cite{kn:pinski} 
were carried out without scalar relativistic effects. Their values
are consequently rather large as compared to ours.
Ruban \etal \cite{kn:rs} have calculated pair potentials for 50$\%$ concentration of Pt using different 
methods and showed that different 
methods give different values of pair potentials. Their nearest neighbor pair-potential is slightly higher
than ours. The effective pair potentials obtained by Pourovskii
\etal \cite{kn:Pourovski} from the neutral charge spheres
GPM method are similar to the estimates of Ruban \etal \cite{kn:rs}.

\vskip 0.2 cm
\n In figure 7 we have shown the ordering energy, anti-phase boundary energy and instability temperatures vs 
concentration of Pt with charge neutral potential parameters including scalar relativistic correction. The ordering energy in all 
three cases Ni$_3$Pt, NiPt and NiPt$_3$ is negative showing the stability of ordered structures
compared to disordered solution. Among all 
three concentrations, NiPt attains maximum value of ordering energy which confirms that $L1_0$ in NiPt system is 
the most stable structure. The anti-phase boundary energy in all these cases Ni$_3$Pt, 
NiPt and NiPt$_3$ comes out to be positive showing 
the ordering structures $L1_2$ for Ni$_3$Pt, $L1_0$ for NiPt and $L1_2$ for NiPt$_3$ as described above. 
The magnitude of instability temperatures 
using the charge neutral potential parameters comes out to be larger than the 
experimental transition temperatures.  However, the  qualitative trend of the change of 
the instability temperatures with changing concentration of Pt is right.
The calculated qualitative phase diagram (instability temperature vs concentration of Pt) shows asymmetric 
feature which is not observed experimentally. This could be due to the neglect
of magnetism in the calculations of effective pair interactions which can have significant effect particularly in the high
concentrations of Ni. Amador \etal \cite{kn:alss} also reported the phase diagram (instability temperature vs concentration of Pt) of 
this system described by the nearest neighbour tetrahedron effective interactions from clusters with appropriate 
effective volume. Their values for transition temperatures are smaller than ours but there is high asymmetry in 
their phase diagram and even the trend is not same as experimental findings.

\section{Conclusion}

\n Our total energy calculations for the ordered alloys indicate that in order to have the correct sign for
the formation energy, it is essential to include relativistic corrections. 
Our analysis of the concentration wave approach indicates that 
for Ni$_3$Pt neither relativistic correction nor the charge transfer effect is essential for 
the correct prediction of the $L1_2$ ground state. For NiPt although scalar relativistic correction is 
not essential, careful treatment of charge transfer effect is a must to predict the correct 
ground state ($L1_0$). For NiPt$_3$ both
these corrections are essential to predict the correct ground state $L1_2$.  
\vskip 0.2cm
\n Although it seems that qualitatively the relativistic corrections and charge transfer effect plays
the essential role
only for the high Pt content alloys, for quantitative prediction of the instability temperature both
these corrections are required across the concentration range. 
\vskip 0.2cm    
\n The main conclusions of this paper are :
\begin{itemize}
\item We have demonstrated that for accurate prediction of the ground
state structures and instability temperatures for alloys with components
with large atomic size differences like NiPt, it is essential to take into
account both relativistic corrections and averaged charge neutrality of the
atomic spheres.
\item We have also demonstrated  the augmented space recursion combined with
the first-principles tight-binding linearized muffin-tin orbitals and the
orbital peeling are both computationally feasible and suitable techniques
for such studies as described above.
\end{itemize}

\n These techniques will form the basis of our further study into similar
alloy systems, but with magnetic effects included.


\begin{figure}
\centering
\epsfxsize=2.5in\epsfysize=3.5in
\rotatebox{270}{\epsfbox{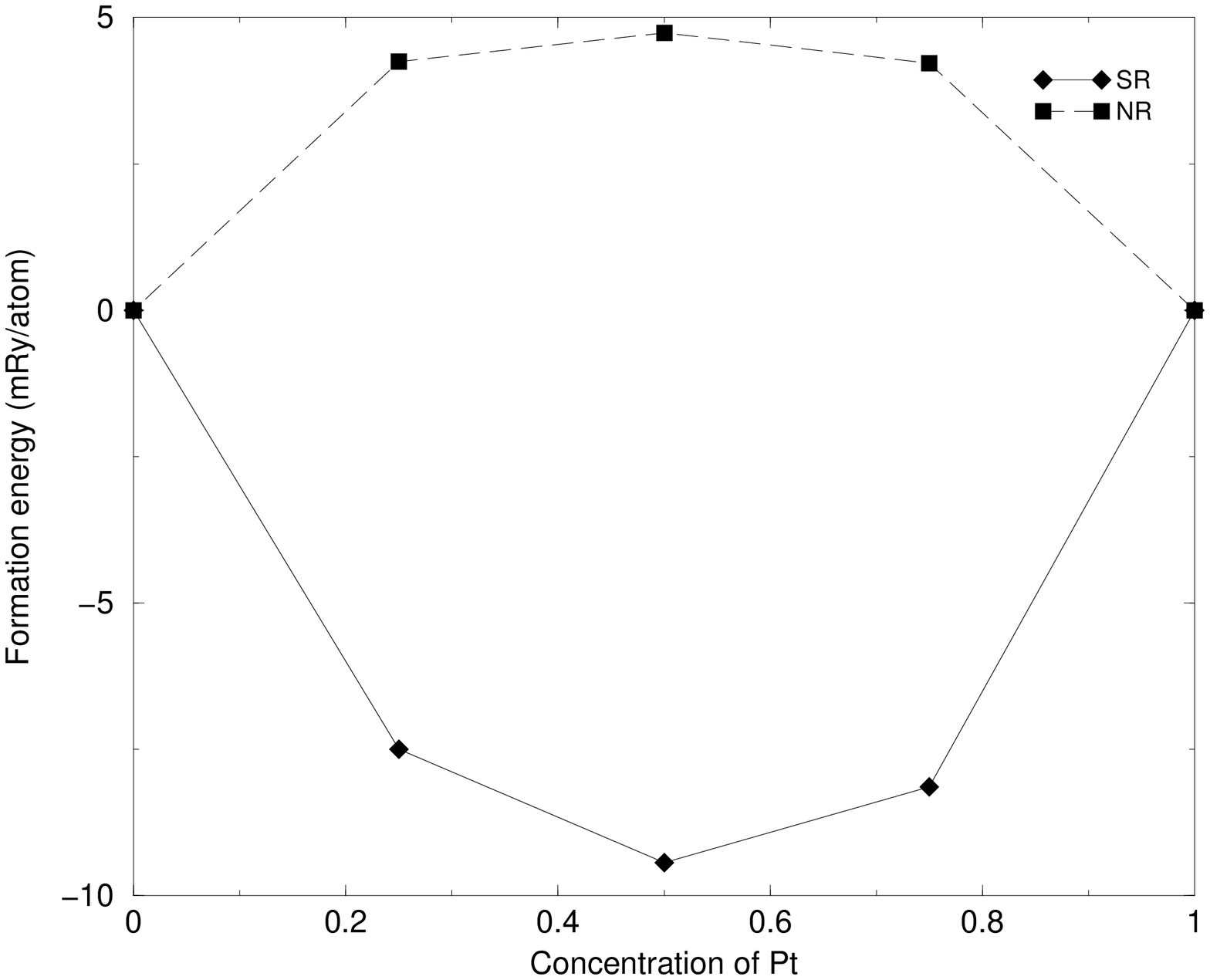}}
\caption{Formation energy vs concentration of Pt with the choice of neutral charge spheres}
\end{figure}

\begin{figure}
\centering
\epsfxsize=2.5in\epsfysize=3.5in
\rotatebox{270}{\epsfbox{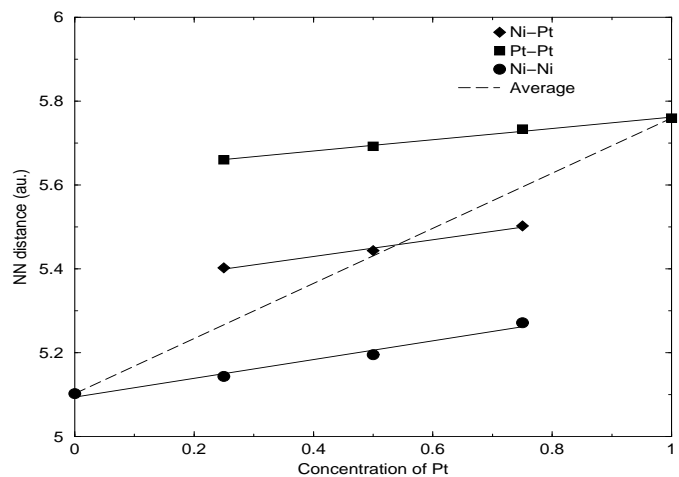}}
\caption{Nearest neighbour distance vs concentration of Pt with the choice of neutral charge spheres. For 
comparison the average bond length given by Vegard's law is shown in dashed line.}
\end{figure}

\begin{figure}
\centering
\epsfxsize=3.5in\epsfysize=5.5in
\rotatebox{0}{\epsfbox{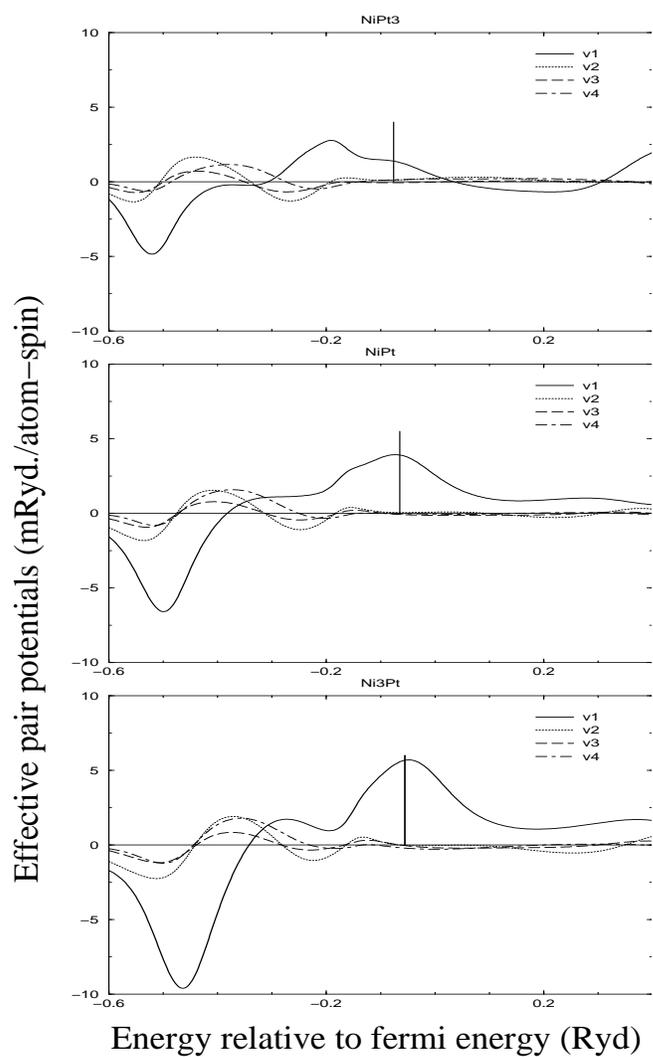}}
\caption{ (i) The effective pair potentials with potential parameters taken from calculations with the choice of 
charged spheres including scalar relativistic corrections.}
\end{figure}

\begin{figure}
\centering
\epsfxsize=6.5in\epsfysize=4.5in
\rotatebox{0}{\epsfbox{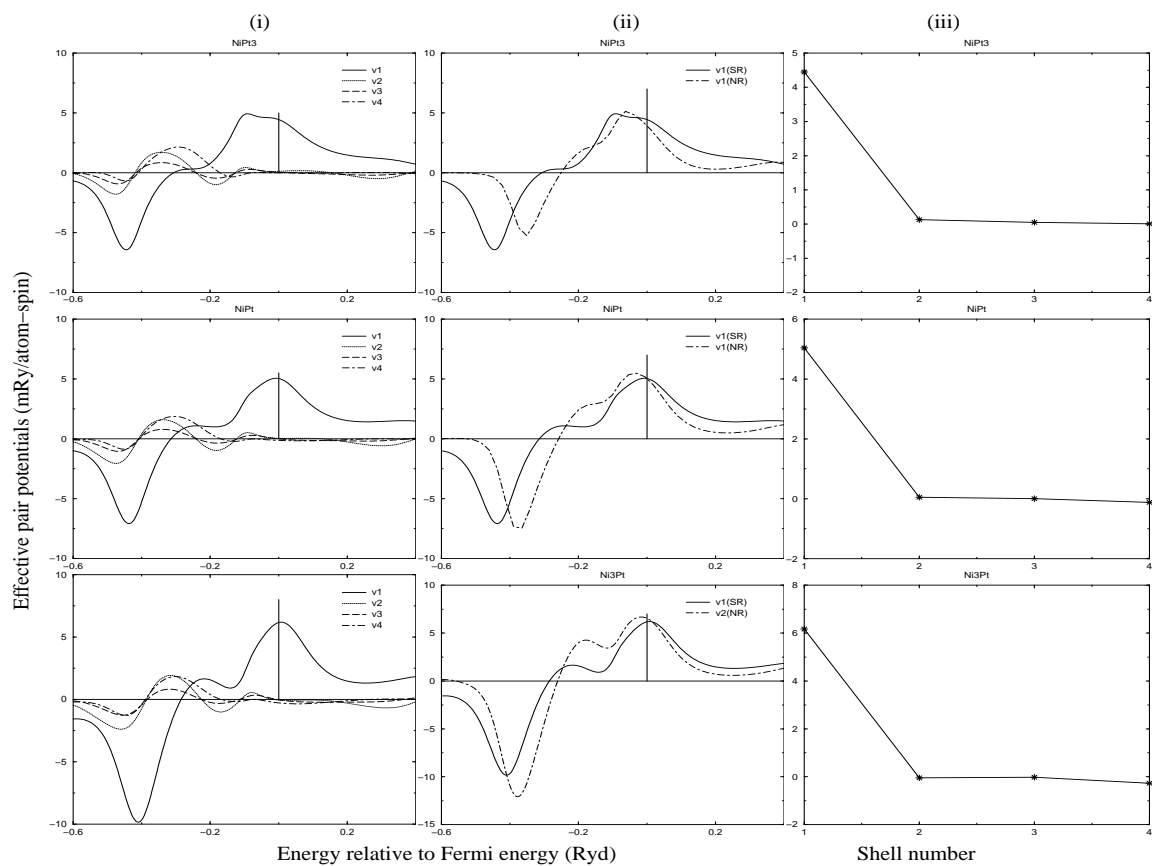}}
\caption{ (i) The effective pair potentials as a function of energy relative to Fermi energy with charge neutral 
potential parameters including scalar relativistic corrections.(ii) Comparison between the first 
nearest neighbour effective pair potentials with scalar relativistic corrections and without scalar 
relativistic corrections by taking charge neutral potential parameters. 
(iii) The effective pair potentials as a function of shell numbers with charge neutral potential parameters 
including scalar relativistic correction.}
\end{figure}

\begin{figure}
\centering
\epsfxsize=5.5in\epsfysize=5.5in
\rotatebox{0}{\epsfbox{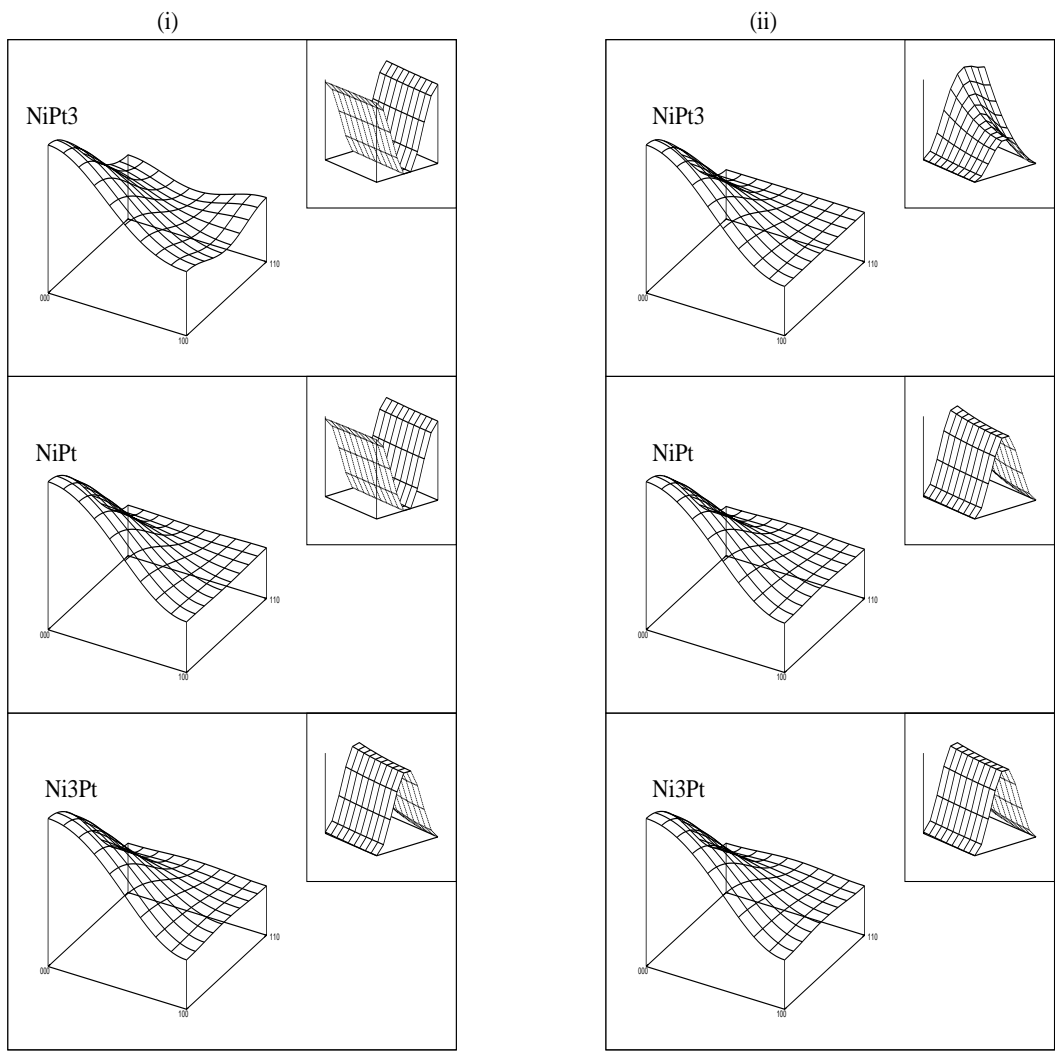}}
\caption{The V ($\vec k$) surface for NiPt alloy system with potential parameters calculated with the choice of 
(i) charged spheres (ii) charge neutral spheres, on $k_z$ = 0 plane. The figures in inset are corresponding 
rescaled V ($\vec k$) surfaces on $k_z$ = 0 plane along the (100) to (110) direction to view the minima.}
\end{figure}

\begin{figure}
\centering
\epsfxsize=6.0in\epsfysize=3.0in
\rotatebox{0}{\epsfbox{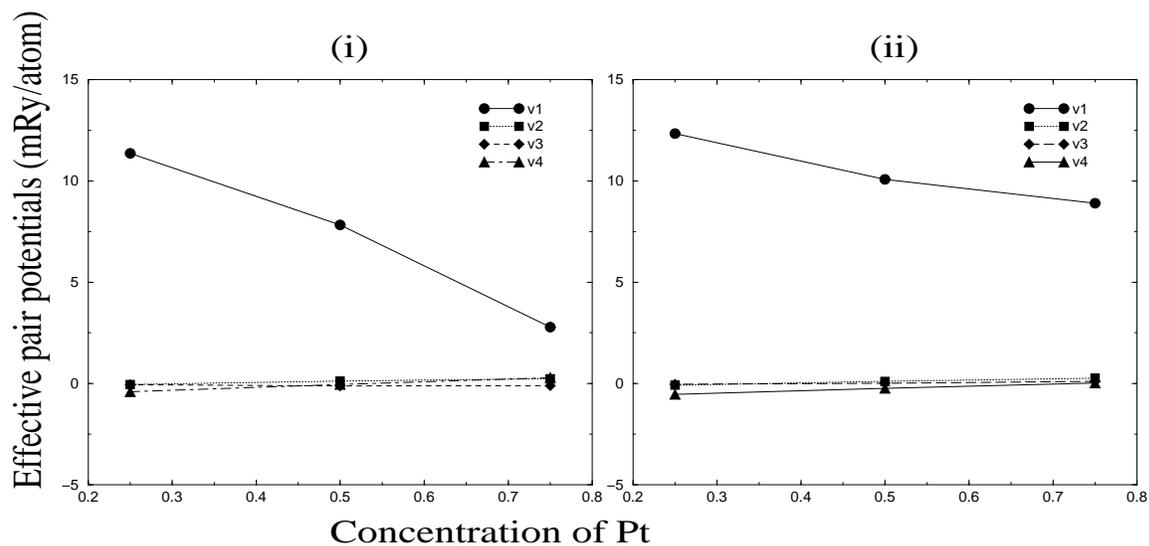}}
\caption{The effective pair potentials vs concentration of Pt with the choice of potential parameters with (i) charged spheres 
including scalar relativistic correction and (ii) charge neutral spheres including scalar relativistic correction.} 
\end{figure}

\begin{figure}
\centering
\epsfxsize=6.0in\epsfysize=2.5 in
\rotatebox{0}{\epsfbox{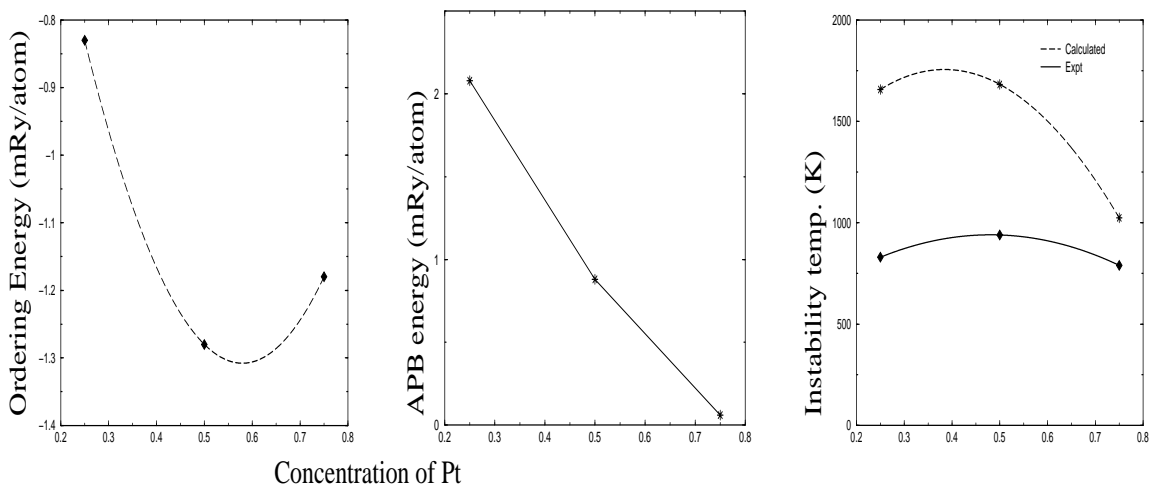}}
\caption{Ordering energy, anti-phase boundary energy and instability temperatures  
vs concentration of Pt with the choice of charge neutral potential parameters including scalar 
relativistic correction.} 
\end{figure}

\end{document}